\documentclass[english,aps,prc,preprint,groupedaddress,amsmath,amssymb]{revtex4}
\usepackage[T1]{fontenc}
\usepackage[latin9]{inputenc}
\usepackage{amstext}

\makeatletter

\providecommand{\tabularnewline}{\\}

\@ifundefined{textcolor}{}
{%
 \definecolor{BLACK}{gray}{0}
 \definecolor{WHITE}{gray}{1}
 \definecolor{RED}{rgb}{1,0,0}
 \definecolor{GREEN}{rgb}{0,1,0}
 \definecolor{BLUE}{rgb}{0,0,1}
 \definecolor{CYAN}{cmyk}{1,0,0,0}
 \definecolor{MAGENTA}{cmyk}{0,1,0,0}
 \definecolor{YELLOW}{cmyk}{0,0,1,0}
 }

\makeatother

\usepackage{babel}
\begin{document}

\title{The Most Negative and Most Positive Expectation Values of the Spin
Operator }

\author{Larry Zamick}

\affiliation{Department of Physics and Astronomy,Rutgers University, Piscataway,New
Jersey,USA 08854}

\altaffiliation{Permanent Address}

\affiliation{Weizmann Institute of Science, Rehovot, Israel}
\begin{abstract}
Formulas for the most positive and most negative values of the expectation
of the spin operator are given and compared with single particle values.
The Nilsson model is used to evaluate these expectations and a scenario
is discussed where the value is greater than one. 
\end{abstract}
\maketitle
The motivation for this work stems from the fact that isoscalar magnetic
moments obtained from mirror pairs and $N=Z$ odd-odd nuclei have
values which are very close to the single $j$ limit--simplicity in
the midst of complexity. We wish to clarify the distinction between
one particle and the many particle aspects of this problem.

For a system of several nucleons we define the expectation value of
the spin operator $\vec{\sigma}=2\vec{S}$

\begin{equation}
\left\langle \sigma\right\rangle =\left\langle \Psi_{J}^{J}\sigma_{z}\Psi_{J}^{J}\right\rangle \label{eq:1}
\end{equation}
 where $\Psi$ is the many-particle wave function in a state with
$M=J$. The magnetic moment of a single nucleon in a state $\psi$$_{j}^{j}$
is called the Schmidt moment. From values of this moment we can infer
that for a single nucleon in a state$[L,1/2]^{j}$ with $j=L+1/2$
the value of $\left\langle \sigma\right\rangle $ is one; for a single
nucleon with $j=L-1/2$ the value is $-j/(j+1)$. We next consider
a system many nucleons and use $LS$ wave functions $[L,S]J$. We
address the problem of what are the most negative and most positive
values of $\left\langle \sigma\right\rangle .$ We find

\begin{equation}
\left\langle \sigma\right\rangle =(1J0J|JJ)\times2S/(1S0S|SS)\sqrt{(2J+1)\times(2S+1)}W(1SJL;SJ)\label{eq:2}
\end{equation}
 where W is a Racah coefficient. We have 
\begin{equation}
(1J0J|JJ)=-\sqrt{J/(J+1)}\qquad\qquad(1S0S|SS)=-\sqrt{S/(S+1)}\label{eq:3}
\end{equation}
 and

\begin{equation}
W=-[S(S+1)+J(J+1)-L(L+1)]/\sqrt{4S(S+1)(2S+1)J(J+1)(2J+1)}\label{eq:4}
\end{equation}
 We find 
\begin{equation}
\left\langle \sigma\right\rangle =[S(S+1)+J(J+1)-L(L+1)]/(J+1).\label{eq:5}
\end{equation}
 Let us consider the extremes. For 
\begin{equation}
J=L-S\qquad\left\langle \sigma\right\rangle =-2SJ/(J+1)\label{eq:6}
\end{equation}
 This is the most negative value this quantity can have for a given
$J$. This expression for several nucleons in $LS$ coupling with
$J=L-S$ is consistent with the expression for a single nucleon with
$j=L-1/2$ $(-j/(j+1))$, as it must be. The maximum value of $\left\langle \sigma\right\rangle $
is obtained by setting $J=L+S$. The value is $2S$. For a single
nucleon the value is one. One can determine $\left\langle \sigma\right\rangle $
from mirror pairs:

\begin{equation}
\left\langle \sigma\right\rangle =(2\mu(IS)-J)/(\mu_{p}+\mu_{n}-1/2)\label{eq:7}
\end{equation}
 where 
\begin{equation}
\mu(IS)=(\mu(T_{z})+\mu(-T_{z}))/2\label{eq:8}
\end{equation}

In a work of Kramer et al.\cite{key-1} the magnetic moment of $^{21}$Mg
is measured, which when combined with the moment of $^{21}$F yields
an isoscalar magnetic moment and an expectation value of the spin
operator. These authors refer to the {}``empirical limits'' .They
use as limits the single particle Schmidt values $-j/(j+1)$ for $j=L-1/2$
and one for $j=L+1/2$ and call the results beyond these limits anomalous.
By this criterion their own value $\left\langle \sigma\right\rangle =1.15(2)$
is anomalous. They also refer to anomalies for $A=9$ found by Matsuta
et al.\cite{key-2} and discussed by Utsuno\cite{key-3}. They obtained
a very large value $\left\langle \sigma\right\rangle =1.44$. A careful
reading of the Matsuta et al.and Kramer et al. papers however shows
that they do not say that these empirical limits are theoretical limits.
Indeed in ref\cite{key-1} the authors report a shell model calculation
with a charge independent interaction which gives a value 1.11, close
to their measured value. They then go on to include a charge symmetry
violating interaction which improves the fit .The final result is
1.15. Their shell model calculation shows that one does not need a
violation of charge symmetry to go beyond the {}``empirical limit''
$\left\langle \sigma\right\rangle =1$.

We would say that their results are not anomalous if the theoretical
limits are used. For $J=5/2$ an $LS$ wave function component with
$L=0$ $S=5/2$ would yield an upper limit of five--much larger than
the Schmidt limit of one. For $L=1$ $S=3/2$ we get three. It would
be correct to say that these configurations are not the major components
of the complete nuclear wave function so it is still surprising that
values greater than one are obtained. For $A=9$ $J=3/2$ there are
several $LS$ configurations with $\left\langle \sigma\right\rangle $
greater than one. For example there is $[311]L=0T=3/2\, S=3/2$ for
which $\left\langle \sigma\right\rangle =3$, and $[221]L=1T=3/2\, S=3/2$
for which $\left\langle \sigma\right\rangle $ is 11/5\cite{key-16}.
Note that the supermultiplet quantum numbers at the left in these
two examples are not needed to evaluate $\left\langle \sigma\right\rangle $-only
$L$ and $S$ are needed. However they are included to show that these
states obey the Pauli principle. Useful references are Wigner\cite{key-16}
and Bohr and Mottelson\cite{key-17}.

The Nilsson one body interaction\cite{key-4} consists of a spin-orbit
term, an $L^{2}$term (to make up for the deficiency of the oscillator
radial shape) and most important a deformed potential $V(r)=1/2m\omega^{2}r^{2}(1-4/3\delta P_{2}(\cos(\vartheta))$.
As the deformation parameter $\delta$ aproaches zero we go towards
the weak deformation limit. For very large $\delta$ we come to the
asymptotic limit where the angle-spin part of the wave function decouples
to the form Y$_{L,\Lambda}$ $\chi$$_{1/2}$,$_{\Sigma}$.For finite
$\delta$ the spin-orbit interaction prevents $\Lambda$ and$\Sigma$
from being good quantum numbers. One gets a sum over various $\Lambda$
and $\Sigma$ with the constraint that $\Lambda+\Sigma=K$. Here the
formula for the laboratory magnetic moment in the rotational model
using the notation of Bohr and Mottelson\cite{key-18}.

\begin{equation}
\mu=g_{R}J+(g_{K}-g_{R})K^{2}/(J+1)(1+\delta_{K,1/2}(2I+1)(-1)^{J+1}b)\label{eq:9}
\end{equation}
 where

\begin{equation}
(g_{K}-g_{R)})b=\left\langle K(g_{L}-g_{R})L_{+}\bar{K}\right\rangle +\left\langle K(g_{S}-g_{R})S_{+}\bar{K}\right\rangle \label{eq:10}
\end{equation}

and $\left.|\bar{K}\right\rangle $ is the time reverse of the state
$\left.|K\right\rangle $. Since $g_{R}$ is $Z/A$ , for mirror pairs
the summed $g_{R}$ is one. Hence , if $K$ is not equal to 1/2 we
obtain

\begin{equation}
2\times\mu(IS)=J+(Kg_{K}-K)\times K/(J+1)
\end{equation}
 where $Kg_{K}=\left\langle g_{L}L_{z}+g_{S}S_{z}\right\rangle $
evaluated in the intrinsic state. Here $g_{L}$ is also one and $g_{S}=2(\mu_{p}+\mu_{n})=1.760$.
Keeping in mind $A=9$ and $A=21$ let us consider intrinsic states
in the weak deformation limit $p_{3/2,K=3/2}$ and $d_{5.2,K=5/2}$
respectively. We find that

\begin{equation}
Kg_{K}-K=(\mu_{p}+\mu_{n}+L-K)\label{eq:11}
\end{equation}

\begin{equation}
2\mu(IS)=J+(\mu_{p}+\mu_{n}+L-K)\times K/(J+1)\label{eq:12}
\end{equation}
 where $Kg_{K}=\left\langle g_{L}L_{z}+g_{S}S_{z}\right\rangle $
is evaluated in the intrinsic state. When we combine this with the
expression at the beginning we obtain

\begin{equation}
j=L+1/2\qquad\qquad2\times\mu(IS,\textrm{Schmidt})=L+\mu_{p}+\mu_{n}\label{eq:15}
\end{equation}

$(\mu(\textrm{Nilsson})-\mu(\textrm{Schmidt}))/\mu(\textrm{Schmidt})=-8.1\%$
for $A=9$; $=-3.8\%$ for $A=21$. Although the percent changes are
rather small the deviations of $\left\langle \sigma\right\rangle $from
unity (The Schmidt value) are large. In more detail

\[
A=9:J=3/2....2\mu(IS)=1.728\qquad\qquad\qquad\qquad\quad\;\left\langle \sigma\right\rangle =0.600
\]

\[
A=21:J=5/2....2\mu(IS)=2.771\qquad\qquad\qquad\qquad\quad\left\langle \sigma\right\rangle =0.713
\]
 Note that for the above states $\left\langle \sigma\right\rangle $
is

equal to one in the intrinsic frame but considerably less than one
in the lab frame.

We now consider $K=1/2$ bands. Since $g_{l}$ and $g_{R}$ are both
one the first term in labeled Eq. \ref{eq:9} vanishes and we have

\begin{equation}
(g_{K}-g_{R})b=\left\langle K|(g_{S}-g_{R})S_{+}\bar{K}\right\rangle \label{eq:16}
\end{equation}

We now list in Table \ref{tab:ExperimentalValuesSpinOperator} experimental
results for $\left\langle \sigma\right\rangle .$ The Schmidt

results are given in Table \ref{tab:Isoscalar-Schmidt-moments}. We
round of all the values to up to three digits beyond the decimal point.

\begin{table}
\caption{\label{tab:ExperimentalValuesSpinOperator}Experimental values of
the spin operator obtained from mirror pairs.}

\centering{}%
\begin{tabular}{c|c|c|c|c|c}
\hline 
Mirror Pairs  & $J$  & Odd Proton  & Odd Neutron  & Sum  & $\left\langle \sigma\right\rangle $\tabularnewline
\hline 
\hline 
$^{9}$Li-$^{9}$C  & 3/2  & 3.439  & -1.394  & 2.048  & 1.434\tabularnewline
\hline 
$^{21}$F-$^{21}$Ne  & 5/2  & 3.93  & -0.983  & 2.947  & 1.176\tabularnewline
\hline 
$^{21}$Ne-$^{21}$Na  & 3/2  & 2.386  & -0.662  & 1.724  & 0.589\tabularnewline
\hline 
$^{23}$Na-$^{23}$Mg  & 3/2  & 2.218  & -0.536  & 1.681  & 0.479\tabularnewline
\hline 
$^{25}$Mg-$^{25}$Al  & 5/2  & 3.646  & -0.855  & 2.790  & 0.766\tabularnewline
\hline 
\end{tabular}
\end{table}

In the single $j$ model for the configurations $(d_{5/2})^{n}$ the
values are as follows:

\[
J=3/2\quad2\mu(IS)=3/5*\mu(IS,Schmidt)=1.728\qquad\qquad\left\langle \sigma\right\rangle =0.6
\]

\[
J=5/2\quad2\mu(IS)=2.880\qquad\qquad\qquad\qquad\qquad\qquad\qquad\left\langle \sigma\right\rangle =1.0
\]

In the weak deformation limit of the Nilsson model one obtains:

\[
J=3/2\qquad\psi_{j,K}=d_{5/2,3/2}\qquad2\mu(IS)=1.637\qquad\qquad\qquad\qquad\left\langle \sigma\right\rangle =0.360
\]

\[
J=5/2\qquad\psi_{j,K}=d_{5/2,5/2}\!\qquad2\mu(IS)=2.771\qquad\qquad\qquad\qquad\left\langle \sigma\right\rangle =0.713
\]

Note that the single $j$ and weak deformation Nilsson values are
not the same. The first two mirror pairs in Table~\ref{tab:ExperimentalValuesSpinOperator}
have isospin $T=3/2$ and the others $T=1/2$. The $T=1/2$ values
of $\left\langle \sigma\right\rangle $ are within the single particle
limits but this is not the case for $T=3/2$. More complete intrinsic
wave functions for the cases where $J=K$ have been obtained by Ripka
and Zamick~\cite{key-5}.They give results for odd proton and odd
neutron nuclei from which we can easily infer the isoscalar results.
The notation in table~\ref{tab:Ripka--Zamick-Expressions} is such
that C(j) is the probability amplitude that the odd particle is in
the jj coupling state n,L,j.

\begin{table}
\caption{\label{tab:Isoscalar-Schmidt-moments}Isoscalar Schmidt moments}

\begin{tabular}{ccc}
\hline 
 & $2\mu$(IS)  & $\left\langle \sigma\right\rangle $\tabularnewline
\hline 
\hline 
s$_{1/2}$  & 0.880  & 1.000\tabularnewline
\hline 
p$_{3/2}$  & 1.880  & 1.000\tabularnewline
\hline 
d$_{5/2}$  & 2.880  & 1.000\tabularnewline
\hline 
p$_{1/2}$  & 0.373  & -1/3\tabularnewline
\hline 
d$_{3/2}$  & 1.272  & -3/5\tabularnewline
\hline 
\end{tabular}
\end{table}

\begin{table}
\caption{\label{tab:Ripka--Zamick-Expressions}Ripka -Zamick Expressions Modified
to Yield Isoscalar Magnetic Moments.}

\centering{}%
\begin{tabular}{cc}
\hline 
p shell  & $2\mu$(IS)\tabularnewline
\hline 
\hline 
J=K=1/2  & 0.3733\tabularnewline
\hline 
J=K=3/2  & 1.7320\tabularnewline
\hline 
s-d shell  & \tabularnewline
\hline 
J=K= 1/2  & 0.1780 C$^{2}$(5/2) -0.1746 C$^{2}$(3/2) +0.3804 C$^{2}$(1/2) -0.5
C(5/2) C(3/2) + 0.5\tabularnewline
\hline 
J=K= 3/2  & 0.1368 {[} C$^{2}$(5/2)-C$^{2}$(3/2){]} -0.3645 C(5/2) C(3/2) +1.5\tabularnewline
\hline 
J=K= 5/2  & 2.7720\tabularnewline
\hline 
\end{tabular}
\end{table}

In Tables \ref{tab:Nilsson-isoscalar-results.} and \ref{tab:Expectation-values-Spin-JK}
we give a selected list of 2$\mu$(IS) and $\left\langle \sigma\right\rangle $.
In Table \ref{tab:Expectation-values-Spin-JK} things are rearranged
to show the evolution of the expectation of the spin operator from
the weak deformation limit to the asymptotic limit.

\begin{table}
\caption{\label{tab:Nilsson-isoscalar-results.}Nilsson isoscalar results.}

\begin{tabular}{ccc}
\hline 
Intrinsic  & $2\mu$(IS)  & $\left\langle \sigma\right\rangle $ \tabularnewline
 State  &  & \tabularnewline
 & \multicolumn{2}{c}{Weak deformation }\tabularnewline
  & \multicolumn{2}{c}{\underline{limit J=j} }\tabularnewline
 p$_{3/2,3/2}$  & 1.728  & 0.600\tabularnewline
 p$_{3/2,1/2}$  & 1.728  & 0.600\tabularnewline
 p$_{1/2,1/2}$  & 0.3733  & -1/3\tabularnewline
 d$_{5/2,5/2}$  & 2.771  & 0.729\tabularnewline
 d$_{5/2,3/2}$  & 2.598  & 0.257\tabularnewline
 d$_{5/2,1/2}$  & 2.706  & 0.543\tabularnewline
 d$_{3/2,3/2}$  & 1.363  & -0.360\tabularnewline
 d$_{3/2,1/2}$  & 1.363  & -0.360\tabularnewline
 s$_{1/2,1/2}$  & 0.880  & 1.000\tabularnewline
 & \multicolumn{2}{c}{\underline{Asymptotic J=3/2} }\tabularnewline
 Y$_{1,1}\uparrow$  & 1.728  & 0.600\tabularnewline
 Y$_{1,1}\downarrow$  & 1.424  & -0.200\tabularnewline
Y$_{1,0}\uparrow$  & 1.880  & 1.000 \tabularnewline
  & \multicolumn{2}{c}{\underline{Asymptotic J=1/2}}\tabularnewline
 Y$_{1,1}\downarrow$  & 0.373  & -1/3\tabularnewline
 Y$_{1,0}\uparrow$  & 0.880  & 1.0\tabularnewline
 & \multicolumn{2}{c}{\underline{Asymptotic J=5/2}}\tabularnewline
 Y$_{2,2}\uparrow$  & 2.771  & 0.729\tabularnewline
 Y$_{2,2}\downarrow$  & 2.446  & -0.143\tabularnewline
 Y$_{2,1}\uparrow$  & 2.609  & 0.286\tabularnewline
 Y$_{2,0}\uparrow$  & 2.880  & 1.000\tabularnewline
  & \multicolumn{2}{c}{\underline{Asymptotic J=3/2}}\tabularnewline
 Y$_{2,2}\downarrow$  & 1.272  & -0.600\tabularnewline
Y$_{2,1}\uparrow$  & 1.728  & 0.600\tabularnewline
 Y$_{2,1}\downarrow$  & 1.424  & -0.200\tabularnewline
 Y$_{2,0}\uparrow$  & 1.728  & 0.600\tabularnewline
  & \multicolumn{2}{c}{\underline{Asymptotic J=1/2}}\tabularnewline
 Y$_{0,0}\uparrow$  & 0.880  & 1.000\tabularnewline
 Y$_{2,1}\downarrow$  & -1/3  & -0.439\tabularnewline
{[}-0.1cm{]} 
 &  & \tabularnewline
\end{tabular}
\end{table}

\begin{table}
\caption{\label{tab:Expectation-values-Spin-JK}Expectation values of the spin
operator for $J=K$.}

\begin{tabular}{cccc}
\hline 
Weak Deformation Limit &  & Asymptotic Limit & \tabularnewline
\hline 
\hline 
  & <$\sigma$> &  & <$\sigma$>\tabularnewline
\hline 
p$_{3/2,1/2}$  & 0.6  & Y$_{1,0}$$\uparrow$  & 1.0\tabularnewline
\hline 
p$_{3/2,3/2}$  & 0.6  & Y$_{1,1}$$\uparrow$  & 0.6\tabularnewline
\hline 
p$_{1/2,1/2}$  & -1/3  & Y$_{1,1}$$\downarrow$  & -0.2\tabularnewline
\hline 
 &  &  & \tabularnewline
\hline 
d$_{5/2,1/2}$  & 0.543  & Y$_{2,0}$$\uparrow$  & 1.000\tabularnewline
\hline 
d$_{5/2,3/2}$  & 0.360  & Y$_{2,1}$$\uparrow$  & 0.286\tabularnewline
\hline 
d$_{5/2,5/2}$  & 0.729  & Y$_{2,2}$$\uparrow$  & 0.729\tabularnewline
\hline 
 &  &  & \tabularnewline
\hline 
d$_{3/2,1/2}$  & -0.360  & Y$_{2,0}$$\uparrow$  & 0.600\tabularnewline
\hline 
d$_{3/2,3/2}$  & -0.360  & Y$_{2,2}$$\downarrow$  & -0.600\tabularnewline
\hline 
 &  &  & \tabularnewline
\hline 
s$_{1/2}$  & 1.000  & Y$_{2,1}$$\downarrow$  & -0.439\tabularnewline
\hline 
\end{tabular}
\end{table}

In the Nilsson model 2 identical particles in the same spacial state
have opposite spins so only the odd particle contributes to $\left\langle \sigma\right\rangle $
and the value is less than or equal to one.To obtain values of $\left\langle \sigma\right\rangle $
greater than one ,components in which the particles are not in the
lowest intrinsic states must be introduced . As an example in the
weak deformation limit we form the intrinsic state where a particle
is promoted from $p_{3/2,3/2}$ to $p_{1/2,-1/2}$. Thus the unpaired
states are $p_{3/2,1/2}$, $p_{3/2,3/2}$and $p_{1/2,-1/2}$. One
obtains 
\begin{equation}
2\mu(IS)=I+K/(I+1)\times[\Sigma(\left\langle L_{z}\right\rangle +1.760\left\langle S_{z}\right\rangle )-K]\label{eq:17}
\end{equation}

This is a $K=3/2$ band and for $J=3/2$ we find that$\left\langle L_{z}\right\rangle =2/3$,
$\left\langle S_{z}\right\rangle =5/6$, 2$\mu$(IS)=1.88 and $\left\langle \sigma\right\rangle =1$.
This does not get us what we want. However if we go to the asymptotic
limit the unpaired states are Y$_{1.0}$$\uparrow$ Y$_{1,1}$$\uparrow$
and Y$_{1,-1}$$\uparrow$. In this limit we find that $\left\langle L_{z}\right\rangle =0$,
$\left\langle S_{z}\right\rangle =3/2,$ $2\mu(IS)=2.164$ and $\left\langle \sigma\right\rangle =1.8$.
This works.

There are many studies of isoscalar magnetic moments. In the work
of Mavromatis et al.\cite{key-6} it is noted that only with a tensor
interaction can one get corrections to the isoscalar momenets of closed
major shells plus or minus one nucleon. The systematics of isoscalar
moments are discussed in the works of Talmi\cite{key-7}, Zamick\cite{key-8},
B.A. Brown\cite{key-9}, Brown and Wildenthal\cite{key-10}, A.Arima\cite{key-11},
I.Towner\cite{key-12}, and I.Talmi\cite{key-13}. Closely related
to mirror pairs are studies of odd-odd $N=Z$ nuclei. It was noted
by Yeager et al.\cite{key-14} that both experimental results and
large scale shell model calculations were close to the single $j$
results. To undersand this corrections to Schmidt in first order perturbation
theory were performed by Zamick et al.\cite{key-15}.They found that
isoscalar corrections were much smaller than isovector ones for $^{57}$Cu
and $^{57}$Ni mirror pairs.The calculations went in the direction
of reducing $\left\langle \sigma\right\rangle .$ For problems other
than this one one will need the supermultiplet quantum numbers of
Wigner\cite{key-16} if one works in the $LS$ coupling basis.

In summary we have shown that the range over which $\left\langle \sigma\right\rangle $
can vary is considerably wider than that given by the single particle
model. We use the Nilsson model to study this problem and we note
some simplicities for the isoscalar mode. The rotational $g$ factor
$g_{R}$ gets replaced by one and the expression for a $K=1/2$ band
simplifies. We show that in this model we can get a value of sigma
greater than one only by allowing more than one nucleon to be unpaired.
In our example we have three unpaired particles.

The author acknowledges a Morris Belkin visiting professorship at
the Weizmann Institute for Spring 2011.He thanks Igal Talmi , Michael
Kirson and Michael Hass for their hospitality; also Justin Farischon
and Diego Torres for considerable help with the manuscript.

\end{document}